\documentclass[a4paper,12pt]{article}
\usepackage{latexsym}
\usepackage{amssymb}
\usepackage{amsmath}
\usepackage[dvips]{graphicx}
\usepackage{subfig}
\begin{document}

\title{Can the ``standard'' unitarized Regge models describe the TOTEM data?}

\author{A. Alkin \thanks{Bogolyubov Institute for Theoretical Physics,
              Metrolologichna 14b, Kiev, UA-03680, Ukraine} \thanks{email: alkin@bitp.kiev.ua} \and
              O. Kovalenko \thanks{Taras Shevchenko Kiev National University - Volodimirska 60, Kiev, UA-03101, Ukraine} \thanks{email: skovalenko@bitp.kiev.ua}
              \and E. Martynov$^{*}$ \thanks{email: martynov@bitp.kiev.ua}}
\maketitle

\begin{abstract}
The standard Regge poles are considered as inputs for two unitarization  methods: eikonal and U-matrix.  It is shown that only models with three input pomerons and two input odderons can describe the high energy data on $pp$ and $\bar pp$ elastic scattering  including the new data from Tevatron and LHC.  However, it seems that  the both considered models (eikonal and $U$-matrix) require a further modification (e.g., to explore nonlinear reggeon trajectories and/or nonexponential vertex functions) for a more satisfactory description of the data at 19.0 GeV$\leq \sqrt{s}\leq$ 7 TeV and 0.01 $\leq |t|\leq $14.2 GeV$^{2}$.
\end{abstract}

\section{Introduction}

The first results of the TOTEM experiment on elastic proton-proton scattering at 7 TeV were published in 2011 \cite{totem1}. Surprisingly none of models (most successful at lower energies and wide intervals of $t$) could  correctly predict the $t$-dependence  of the $pp$ differential cross sections at LHC energy \cite{totem2, totem3}.

The data on elastic $pp$ and $\bar pp$ scattering at energies up to Tevatron one were quite successfully described in three types of models. The fist type deal with the n intercept larger than one pomerons (e.g. Donnachie-Landshoff model, see  \cite{DLND} and the recent papers \cite{DL-2011}, \cite{LengyelTarics}). However such models have defect: they violate the Froissart bound on the total cross sections and have to be unitarized even if they are in agreement with the data. This fearture give rises to various methods of unitarization of the input amplitudes violating the unitarity. The most known models of such type are the eikonal (see a general discussion on the eikonal method in \cite{Collins}) (or quasi-eikonal \cite{qeik}) and $U$-matrix \cite{UM} (or its generalization \cite{qUM}). The models of the third type are constructed taking into account restrictions of amplitude analyticity and unitarity from the beginning.

The TOTEM data have been analyzed in a few recent publications. Here, for the sake of shortness we do not review the obtained results.  The corresponding references and discussion on various models and its agreement or disagreement with the old data and recent D0 \cite{D0} and TOTEM \cite{totem2} ones are well presented in \cite{Godizov}  and in the comprehensive review \cite{Dremin} .

We will concentrated here on the eikonal  and $U$-matrix models in their "standard" form with the simplest input reggeon amplitude. The name "standard" refers to employing  reggeons with linear trajectories and exponential form of vertex functions. Our aim is to check if such models are able to describe data on $pp$ and $\bar pp$ elastic scattering at high energies and in all available transferred momenta including the new data . We will also to determine the minimal number of the input reggeons that contribute to amplitude and provide a reasonable agreement with the data.

\section{``Standard'' schemes of unitarization}
\subsection{Impact amplitudes}
Let define the elastic amplitude in impact parameter
(two-dimensional vector $\vec b$) representation, or in short impact amplitude, as
\begin{equation}\label{eq:impact ampl}
H(s,b)=\frac{1}{4s}\int \frac{d^{2}\vec k}{(2\pi)^{2}}e^{i\vec k\vec b}A(s,t)=
 \frac{1}{8\pi s}\int \limits_{0}^{\infty} dk k J_{0}(b\sqrt{-t})A(s,t)
\end{equation}
with the inverse transformation
\begin{equation}\label{eq:impact inv}
A(s,t)=16\pi^{2}s\int \frac{d^{2}\vec b}{(2\pi)^{2}}e^{-i\vec k\vec b}H(s,b)=
 8\pi s\int \limits_{0}^{\infty} db b J_{0}(b\sqrt{-t})H(s,b)
\end{equation}
where $A(s,t)$ is an elastic amplitude and $J_{0}(z)$ is the Bessel function, $\vec k$ is two-dimensional vector and  $t\approx -k^{2}_{\bot}$.

One can obtain from the unitarity equation for $A(s,t)$ (see e.g. \cite{Collins}) the unitarity equation for
$H(s,b)$ which in a  general form can be written as
\begin{equation}\label{eq:u.c.impact gen}
\Im mH(s,b)=|H(s,b)|^{2}+G_{inel}(s,b),
\end{equation}
where $G_{inel}(s,b)>0$  takes into account inelastic
intermediate states from the original unitarity condition for $A(s,t)$. Therefore
\begin{equation}\label{eq:impact bound-1}
\Im mH(s,b)>0.
\end{equation}
and from the unitarity one can obtain
\begin{equation}\label{eq:impact bound-2}
|H(s,b)|\leq 1.
\end{equation}
For the observed cross sections one can derive the following expressions
\begin{equation}\label{eq:sigt}
\sigma_{t}(s)=\frac{1}{s}\Im mA(s,0)=8\pi\int\limits_{0}^{\infty}db
b \Im mH(s,b) 
\end{equation}
\begin{equation}\label{eq:sigel}
\sigma_{el}(s)=\frac{1}{16\pi s^{2}}\int\limits_{-\infty}^{0}dt
|A(s,t)|^{2}=8\pi \int\limits_{0}^{\infty} db
b|H(s,b)|^{2} 
\end{equation}
\begin{equation}\label{eq:siginel}
\sigma_{inel}(s)=8\pi \int\limits_{0}^{\infty} db
b (\Im m H(s,b)-|H(s,b)|^{2}).
\end{equation}

\subsection{Output amplitudes}
\begin{figure}[h]
\begin{center}
\includegraphics[scale=0.4]{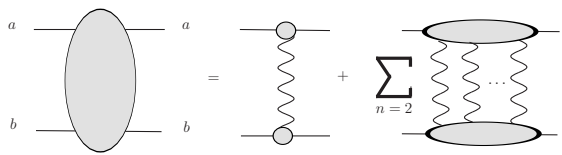}
\caption{Multipomeron contributions to elastic scattering amplitude}
\label{fig.1}
\end{center}
\end{figure}

We consider two known methods for unitarisation of input amplitudes $h^{ap}(s,b)$ where $a=p$ or $a=\bar p$.
Both of them can be considered as the particular cases of sum of multireggeon exchanges (see fig.~\ref{fig.1}). Just for illustration we show here only pomeron contributions, a more realistic case will be considered later on.
\begin{figure}[h]
\begin{center}
\includegraphics[scale=0.4]{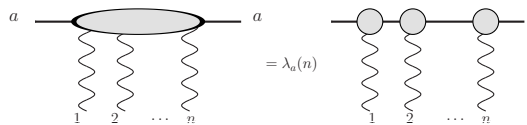}
\caption{Amplitude of interaction of two hadrons with $n$ pomerons in a pole approximation but with phenomenological factor $\lambda_{a}(n)$.}
\label{fig.2}
\end{center}
\end{figure}

Assuming that the  two-hadrons--{\it n}-pomeron amplitude is proportional to the product of the two-hadron--pomeron vertices depending only on $n$ (it is a pole approximation for intermediate states) as shown in fig.~\ref{fig.2}, one can obtain
\begin{equation}\label{eq:Hsb}
H(s,b)=\frac{1}{2i}\sum\limits_{n=1}^{\infty}\frac{\lambda_{a}(n)\lambda_{b}(n)}{n!}[2ih(s,b)]^{n}.
\end{equation}

Moreover, assuming either $\lambda (n)=\lambda^{n}$ or $\lambda (n)=\lambda^{n}n!$, where $\lambda =\lambda_{a}\lambda_{p} (=\lambda_{ap})$, we obtain from Eq.~(\ref{eq:Hsb}) two well-known schemes of pomeron unitarization: eikonal \cite{Collins,3Pom1,3Pom2}\ or quasi-eikonal \cite{qeik}  and $U$-matrix or quasi-$U$-matrix models \cite{UM,qUM}.

\noindent
{\it Eikonal and Quasi-eikonal unitarisation}
\begin{equation} \label{eq:QEik}
 H^{ap}(s,b)=\frac{ e^{2i\lambda_{ap}h^{ap}(s,b)}-1}{2i\lambda_{ap}}, \quad \lambda_{ap}(n)=(\lambda_{ap})^{n}
\end{equation}
If $\lambda_{ap}=1$ then Eq.~(\ref{eq:QEik}) describes the pure eikonal unitarization.

\noindent
{\it U-matrix and quasi-U-matrix unitarization}
\begin{equation}\label{eq:QUM}
 H^{ap}(s,b)=\frac{h^{ap}(s,b)}{1-2i\lambda_{ap}h^{ap}(s,b)} \quad \lambda_{ap}(n)=(\lambda_{ap})^{n}n!.
\end{equation}
For $\lambda_{ap}=1/2$ we have the pure U-matrix unitarization.

In the both cases
\begin{equation}\label{eq:lambd-eik}
\lambda_{ap}\geq 1/2
\end{equation}
because full amplitude must satisfy the inequality $|H^{ab}|\leq 1$.

\subsection{Standard Regge pole input amplitudes}
For input amplitudes we consider the standard Regge pole contributions
\begin{equation}\label{eq:amp-in}
a^{\bar pp}_{pp}(s,t)=\sum _{R_{+}}a_{+}(s,t)\pm \sum_{R_{-}}a_{-}(s,t)
\end{equation}
At high energy ($s\gg m_{p}^{2}$)  the input amplitudes can be written as following
\begin{equation}\label{eq:ampas-evodd}
a_{\pm}(s,t)=\binom{-1}{i}g_{R_{\pm}}(t)(-i\tilde s)^{\alpha_{R_{\pm}}(t)},
\end{equation}
with $\tilde s=s/s_0 $, $s_0$=1GeV$^2$.

There are two kinds of Regge poles contributing to amplitudes. The first ones are poles with intercepts close  to 1 (or large than 1). They are crossing even pomeron (or pomerons) and crossing odd odderon (or odderons). The second kind of contributions are so called secondary reggeons with intercepts less than 1. They are $f, a_2, \omega, \rho$ with $\alpha (0)\sim 0.5$ and other with lower intercepts.  For high energy $pp (p\bar p)$ scattering it is sufficient to consider in the amplitudes one effective crossing even secondary reggeon  ($R_{+}$) and one effective crossing odd secondary  reggeon ($R_{-}$).

In what following we keep the arguments of the \cite{3Pom1} and consider the secondary crossing even  and crossing odd reggeons with the fixed intercepts and slopes (strictly speaking, effective trajectories are not coincided with those for $f$- and $\omega$- reggeons while at high energy their contributions to amplitudes are closed)
\begin{equation}
\begin{array}{ll}
& \alpha_{+}(0)=0.69, \qquad \alpha'_{+}=0.84 \mbox{GeV}^{-2}\\
& \alpha_{-}(0)=0.47, \qquad  \alpha'_{-}=0.93 \mbox{GeV}^{-2}.
\end{array}
\end{equation}

\section{One (two) pomeron and one odderon model\\ (1P(2P)+O-model)}

{\it A. One exponent form of the vertex functions $g_{Pi}(t), i=1, 2$  and $g_{O}(t)$}.

For eikonal form of the amplitude (\ref{eq:QUM}) one and two pomeron models with the exponential vertex functions ($g_{i}(t)=g_{i}\exp(B_{i}t)$) input amplitudes have the form
\begin{equation}\label{eq:s-1P-1Oexp}
a_{i}(s,t)=\eta_{i}g_{i}\exp(B_{i}t)(-i\tilde s)^{\alpha_{i}(0)+\alpha^{'}_{i}t}
\end{equation}
where $i=P_1, P_2, O, +,-$ and $\eta_i=-1$ for pomerons and crossing even secondary reggeon,  $\eta=i$ for odderon and crossing odd reggeon.
In accordance with Eq.~(\ref{eq:impact ampl}) the impact amplitudes $h_i(s,b)$   are  the following
\begin{equation}\label{eq:input-h}
h_{i}(s,b)=\eta_i\frac{g_{i}}{16\pi sr_{i}^2}(-i\tilde s)^{\alpha_{i}(0)}\exp(-b^2/4r_{i}^2)
\end{equation}
where
\begin{equation}\label{eq:r2}
r_{i}^2=B_{i}+\alpha^{'}_{i}\ln(-i\tilde s),\qquad   i= P_1, P_2, O, +,-.
\end{equation}
These models have been considered by  Petrov and Prokudin \cite{3Pom1}. They found that the P+O model  as well as 2P+O one, the both with linear trajectories and with exponential form of the vertex functions $g(t)$ (for all reggeons including pomerons and odderon), failed to describe data with a sufficient  quality.

{\it B. Two exponents form of the vertex functions $g_{Pi}(t), i=1, 2$ and $ g_{O}(t)$}.
To check if a more complicated vertex function $g(t)$ can improve the data description we have considered for pomeron and odderon terms (or for some of them) a combination of two exponents with the different slopes $B$. Namely, we write for $i=P_1, P_2, O$
\begin{equation}\label{eq:twoexps}
g_{i}(t)=g_{i}[c_{i}\exp(B_{i1}t)+(1-c_{i})\exp(B_{i2}t)].
\end{equation}
Changes for the impact amplitudes $h_{Pi}(s,b)$ and $h_{O}(s,b)$ are quite evidently follow from the Eqs.(\ref{eq:input-h},\ref{eq:twoexps}).

We do not give here the results of the fit for this modified model, we rather note that such an extension of the model in spite of some its improvement does not lead to a satisfactory description of  the TOTEM data.

\section{Three pomeron and one odderon model\\
(3P+O-model)}
This model has been suggested and considered  by V. Petrov and A. Prokudin in \cite{3Pom1}. The input amplitudes $h^{ab}(s,b)$ in the model are written as contribution of three pomerons, one odderon and two secondary (even and odd) reggeons
\begin{equation}\label{eq:3P+O-model}
\begin{array}{ll}
 h^{\bar pp}_{pp}(s,b)&=h_{P_{1}}(s,b)+h_{P_{2}}(s,b)+h_{P_{3}}(s,b)+h_{+}(s,b)\\
 & \pm \big (h_{O}(s,b)+h_{-}(s,b) \big)
\end{array}
\end{equation}
where input impact amplitudes are given by Eq.~(\ref{eq:input-h}). Then the eikonalized amplitudes $A^{\bar pp}_{pp}(s,t)$  are defined by Eq.~(\ref{eq:QEik}) with $\lambda^{ap}=1$ and Eq.~(\ref{eq:impact inv}).
They have obtained very good description of the $pp$ and $\bar pp$ data for the total cross sections $\sigma_{tot}(s)$, the ratios of  real to imaginary part of the forward elastic scattering amplitudes $\rho(s)$ at $\sqrt{s}\ge 8$ GeV and for the differential cross sections $d\sigma(s,t)/dt$ at $\sqrt{s}\geq$ 23 GeV and 0.01 $\leq |t|\leq $ 14 GeV$^2$ (this data set contains totally 961 points).

 It is necessary to note here that our data set  differs of those used in \cite{3Pom1,3Pom2}. Our data set  has been proposed by J.R.~Cudell, A.~Lengyel and E.~Martynov (CLM) \cite{CLM}, who built a coherent set of all existing data for $4\le \sqrt{s} \le 1800$ GeV and $0\le\vert t\vert\le 15$ GeV$^2$ \cite{CLM} which can be found
 in the HEP DATA system \cite{HEPdata,DurhamDB}. CLM have performed a detailed study of the systematic errors and gathered into a common format more than 260 subsets of data from more than 80 experimental papers. We suggest to use the CLM data set as a \textit{standard dataset}. The latest (with some corrections) updated version of it  including TOTEM \cite{totem1,totem2,totem3} and D0 \cite{D0} data is available online \cite{CLMdata}.
 The set used for the given analysis contains 1723 points In the region described above. There are only 31 points from the 3 groups, 8 points at $\sqrt{s}=$26.946 GeV, 11 points at 30.7 GeV and 12 points at 53.018 GeV  for  $d\sigma_{pp}/dt$ were excluded from the all data presented in \cite{DurhamDB} because these groups are strongly deviated from the rest data points and can slightly distort the fit.

\begin{figure}[h]
\centering
\subfloat[fig][\small{from the \cite{3Pom1}}]{
 \includegraphics[scale=0.7]{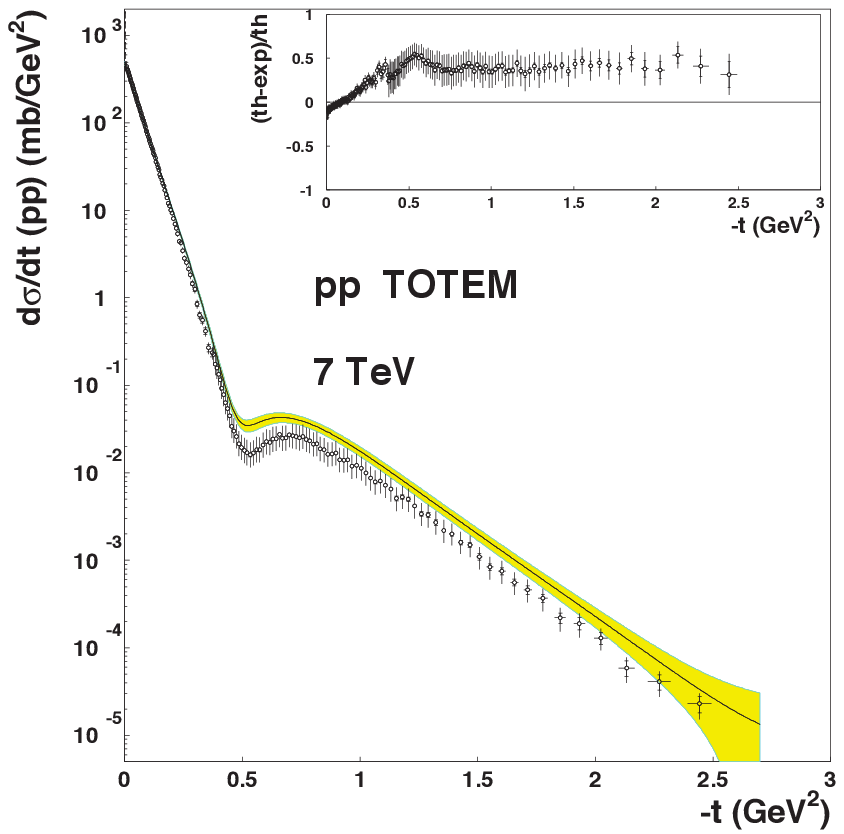}
 \label{fig3a}
}
\qquad
\subfloat[fig][\small{after refitting}]{
 \includegraphics[scale=0.35]{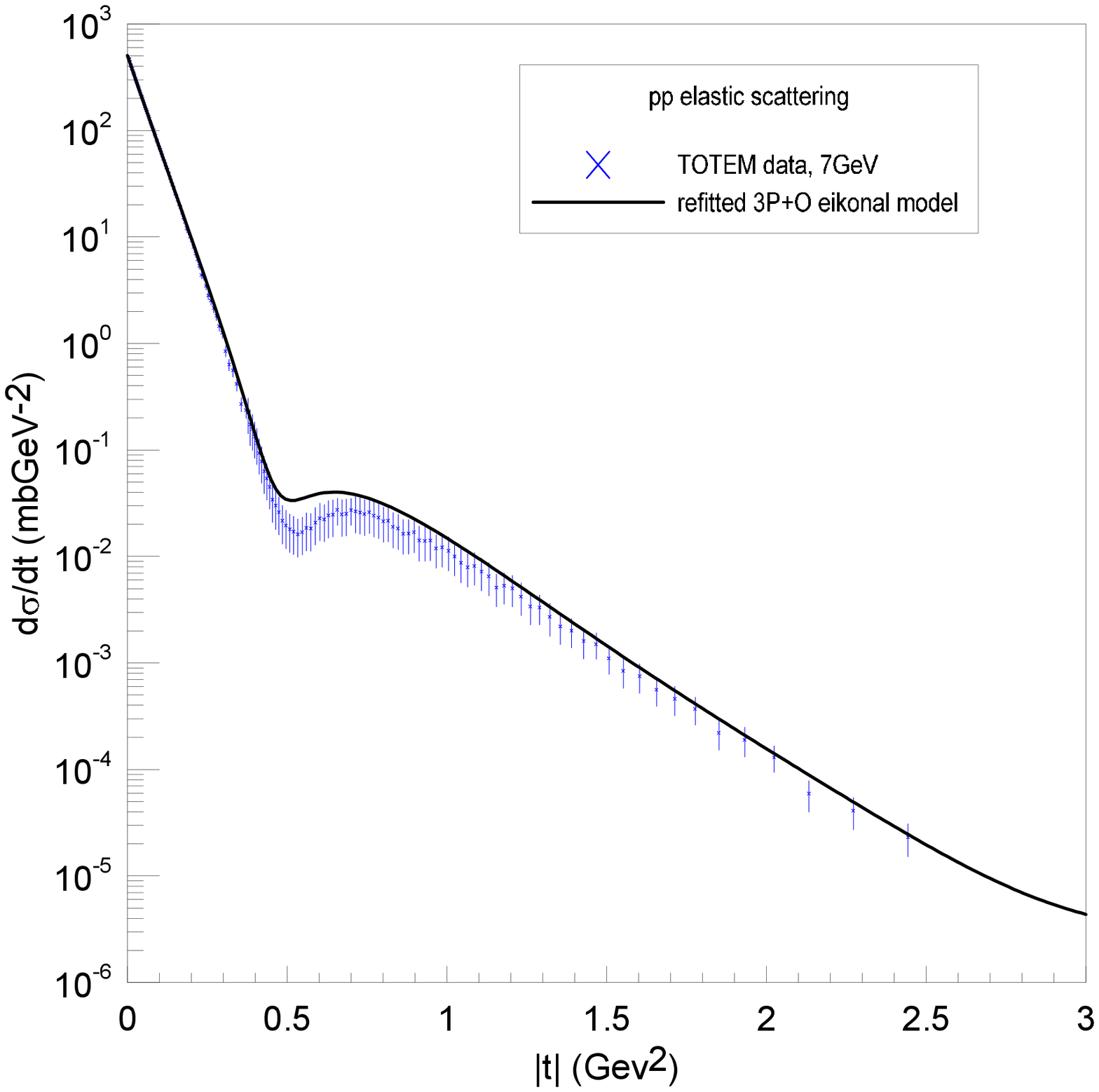}
 \label{fig3b}
}
\caption{Description of the TOTEM data in the three pomerons and one odderon model \cite{3Pom1}}
\label{fig3}
\end{figure}

 As was shown in the paper \cite{totem1} where the first data at LHC energy 7 TeV being presented, the model \cite{3Pom1} does not describe the TOTEM data. This statement is confirmed in the recent paper \cite{3Pom2}.by authors of the model   (we reproduce Fig. 4 of their paper in the fig.~\ref{fig3a}). We have performed refit of the model with our data set including the TOTEM data and found that the resulting theoretical curve for $d\sigma/dt$ at 7 TeV is still out of the data  at $|t|>0.4 $ GeV$^{2}$ as shown on the fig.~\ref{fig3b}. Thus we can conclude that the 3P+O-model should be modified either adding new terms or using nonlinear pomeron (odderon) trajectories and more complicated non exponential vertex functions. In what follows we consider the first possibility, namely we would like to check if an adding the second odderon term can help to improve agreement of the model with new data.

\section{Three pomerons and two odderons model \\
(3P+2O-model)}
In this section we consider the minimal extention of the 3P+O-model (\ref{eq:3P+O-model}) adding the second odderon term.
\begin{equation}\label{eq:3P+2O-model}
\begin{array}{ll}
& h^{\bar pp}_{pp}(s,b)\!\!=\!\!h_{P}(s,b)+h_{+}(s,b)\pm\big ( h_{O}(s,b)+h_{-}(s,b)\big), \\
& h_{P}(s,b)=\sum\limits_{i=1}^{3}h_{P_{i}}(s,b), \quad
h_{O}(s,b)=\sum\limits _{i=1}^{2}h_{O_i}(s,b)
\end{array}
\end{equation}
where each input term is still  in a "standard"  form (\ref{eq:ampas-evodd}),(\ref{eq:s-1P-1Oexp}) with linear trajectory $\alpha_{i}=\alpha_{i}(0)+\alpha'_{i}t$ and exponential vertices $g_{i}(t)=g_{i}\exp(B_{i}t)$.

Later we will check another possibilities to improve the eikonal and U-matrix models (nonlinear trajectories, non exponential vertices and so on).

\begin{figure}[h]
\centering
\subfloat[fig][\small{Total $pp$,  $\bar pp$ cross sections.}]{
 \includegraphics[scale=0.35]{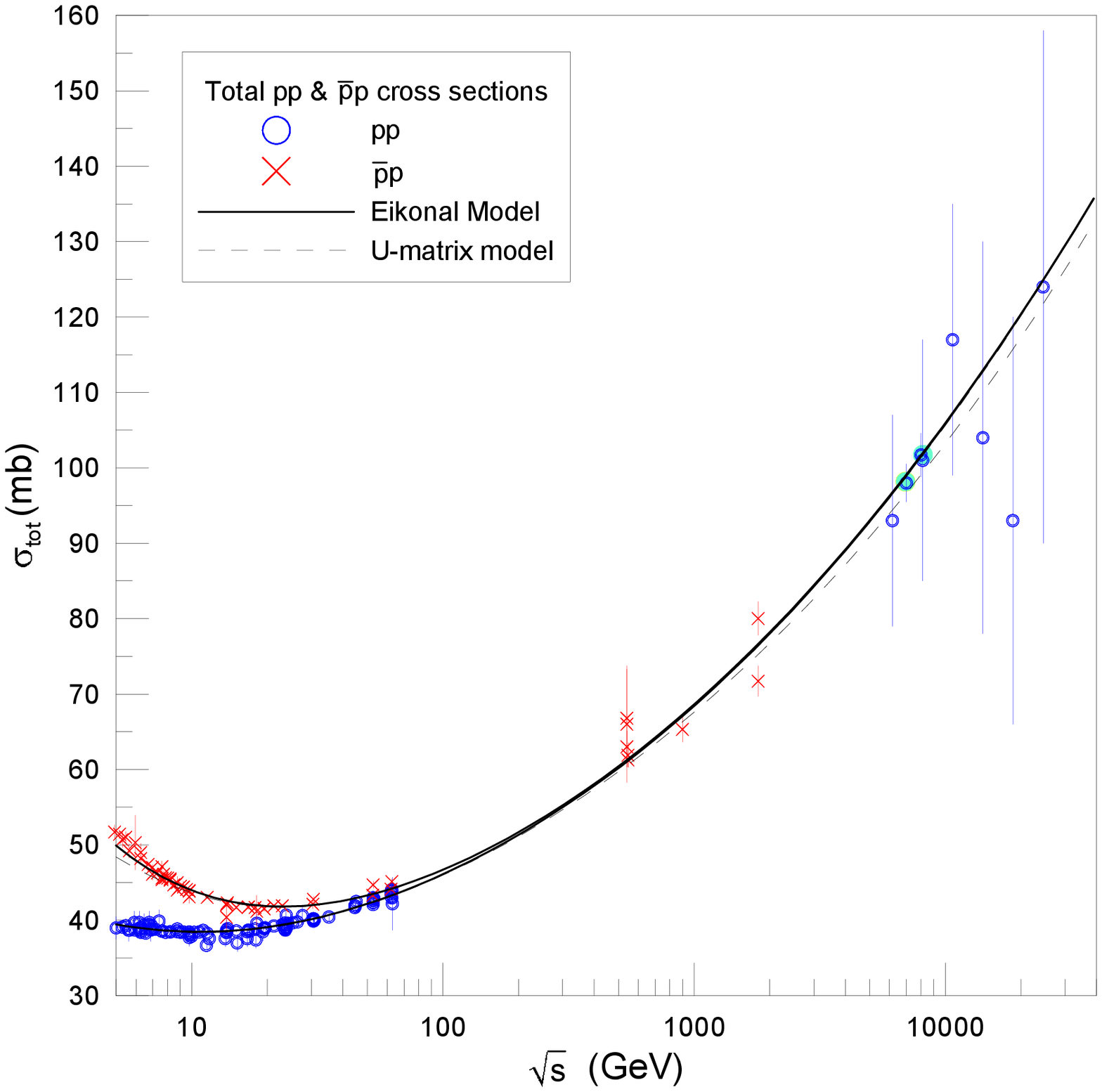}
 \label{fig4a}
}
\qquad
\subfloat[fig][\small{Total, elastic and inelastic $pp$ cross sections.}]{
 \includegraphics[scale=0.35]{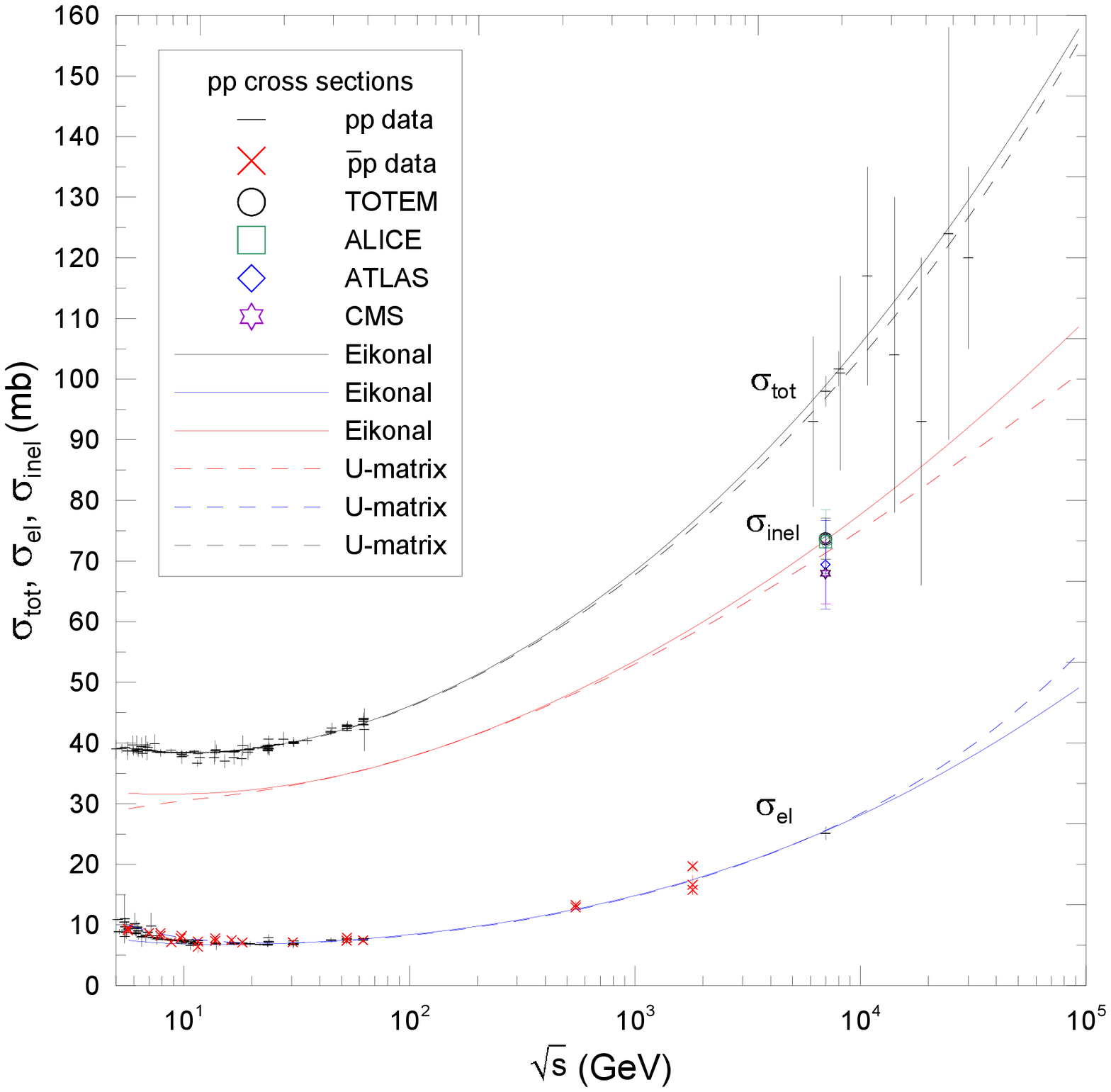}
 \label{fig4b}
}
\caption{Total, elastic and inelastic cross sections in 3P+2O models. Solid (dashed) line shows the eikonal ($U$-matrix) results.}
\label{fig4}
\end{figure}

We considered the both methods of unitarization, eikonal one, Eq.~(\ref{eq:QEik}), with $\lambda^{ap}=1$ and $U$-matrix one, Eq.~(\ref{eq:QUM}), with $\lambda^{ap}=1/2$.
Fit has been performed with the data in the region
\begin{equation}
\begin{array}{lll}
& \rm{for} \quad  \sigma_{tot}(s) \quad  \rm{and}\quad  \rho(s) & \quad \rm{at} \quad \sqrt{s}\geq 5 \rm{GeV},\\
& \rm{for} \quad d\sigma(s,t)/dt & \quad  \rm{at} \quad  \sqrt{s}\geq 19.0\rm{GeV}\\
& &  0.01\leq |t|\leq 14.2\rm{GeV}^{{2}}  
\end{array}
\end{equation}
The resulting  description of the data is shown in the fig.~\ref{fig4}--fig.~\ref{fig7}. Parameters of the models are given in the Table~\ref{tab1}.
\begin{figure}[h!]
\begin{center}
\includegraphics[scale=0.35]{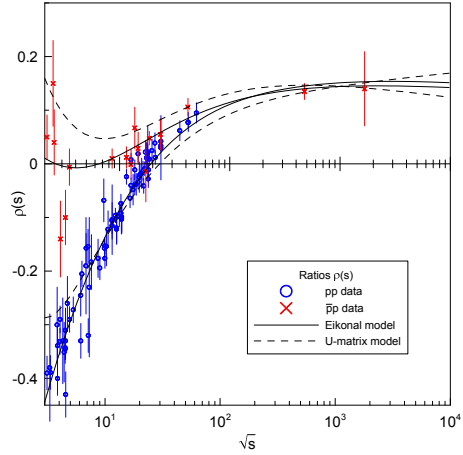}
\caption{The ratio of the real to imaginary part of the forward scattering amplitude. Solid (dashed) line shows the eikonal ($U$-matrix) model results.}
\label{fig5}
\end{center}
\end{figure}
\begin{figure}[h!]
\centering
\subfloat[fig][\small{Differential elastic $pp$ cross sections.}]{
 \includegraphics[scale=0.35]{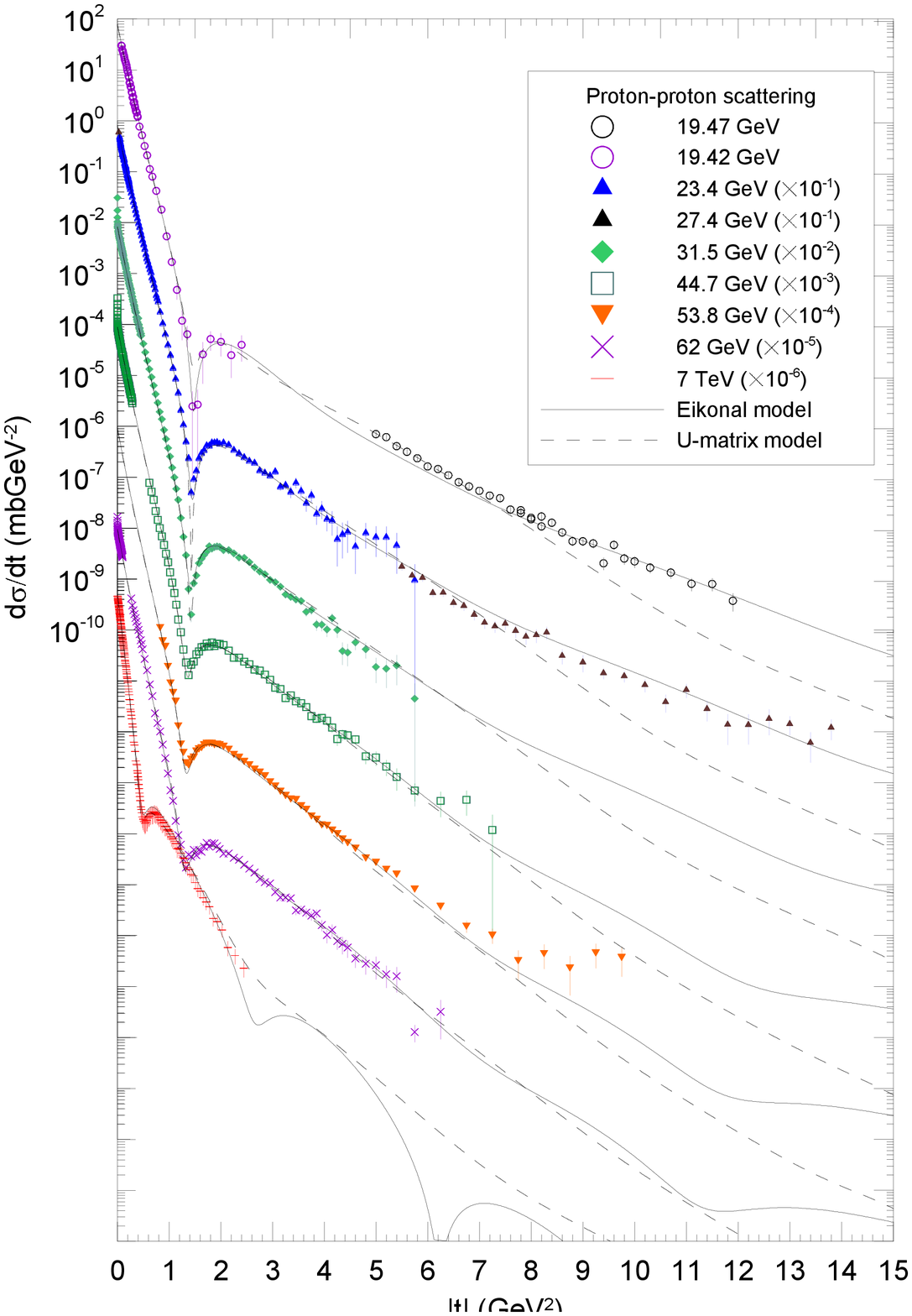}
 \label{fig6a}
}
\qquad
\subfloat[fig][\small{Differential  elastic $\bar pp$  cross sections.}]{
 \includegraphics[scale=0.35]{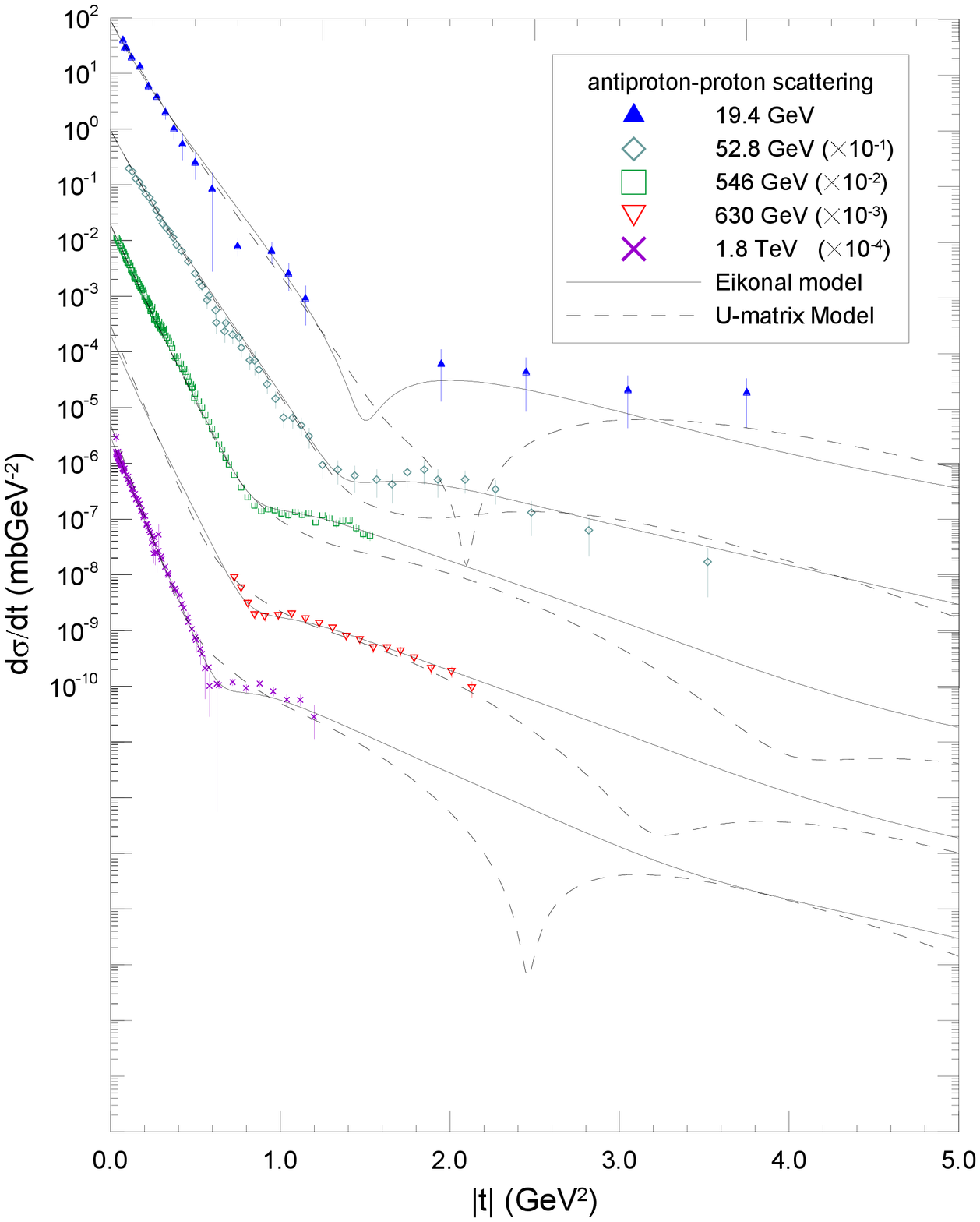}
 \label{fig6b}
}
\caption{Elastic $pp$ and $\bar pp$ scattering in the 3P+2O model. Solid (dashed) line shows the eikonal ($U$-matrix) results.}
\label{fig6}
\end{figure}

\begin{figure}[h!]
\centering
\subfloat[fig][\small{Prediction of the models for higher $|t|$.}]{
 \includegraphics[scale=0.35]{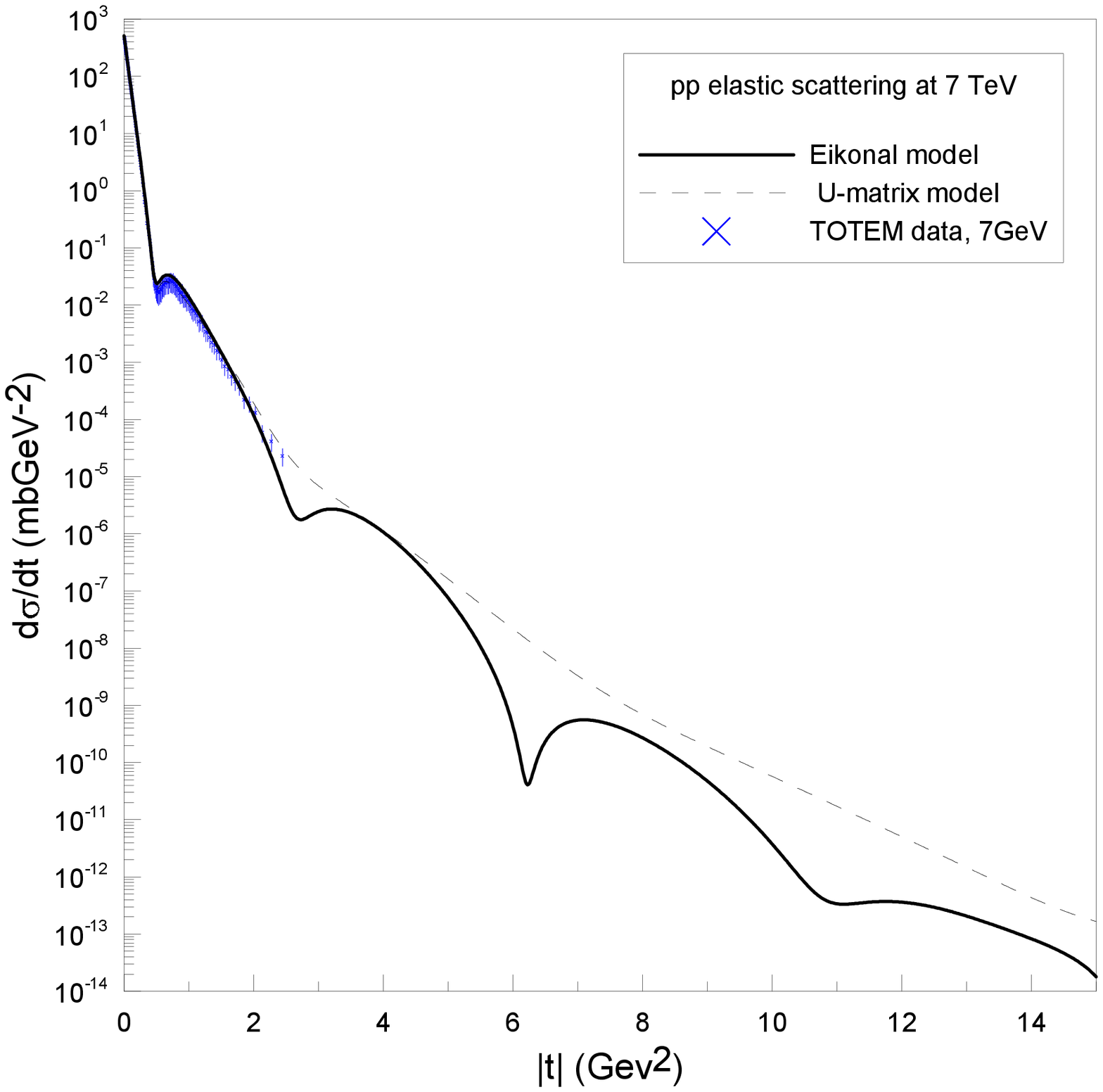}
 \label{fig7a}
}
\qquad
\subfloat[fig][\small{Description of the data}]{
 \includegraphics[scale=0.35]{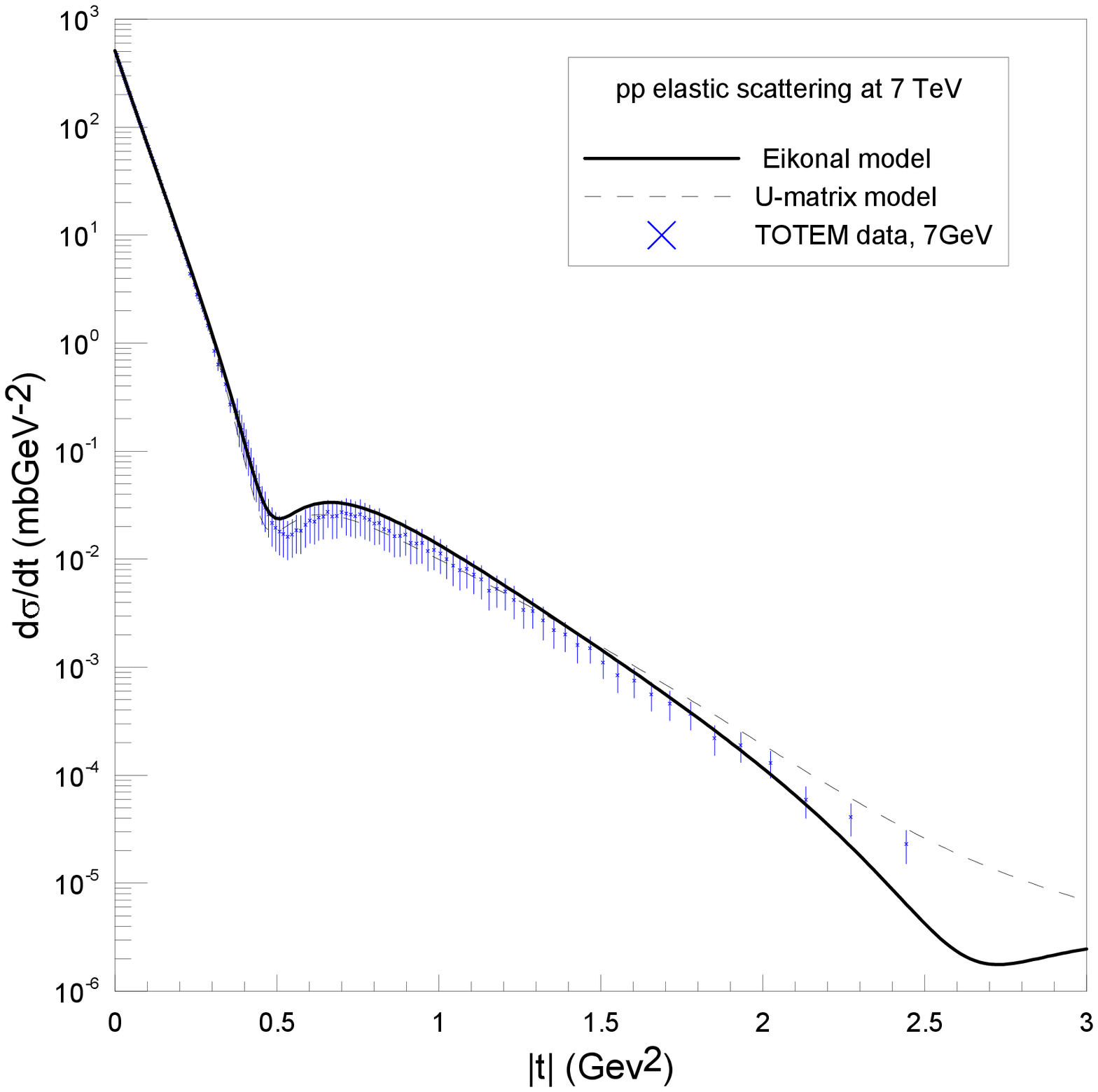}
 \label{fig7b}
}
\caption{Description of the TOTEM data and predictions for high $|t|$ of the 3P+2O models. Solid (dashed) line shows the eikonal ($U$-matrix) results.}
\label{fig7}
\end{figure}

Let us briefly comment the results.

Addition of the second odderon term leads to certain improvement of  the data description.

Agreement with the TOTEM data is essentially better but some defects are appeared at lower energies, especially at large $|t|$ and for $\bar pp$ differential cross
sections (see figs.~\ref{fig6} and \ref{fig7}).

Big difference for $\rho_{pp}$ and $\rho_{\bar pp}$ (shown in fig.~\ref{fig5}) is disappeared at very high energy. In the eikonal model the difference is negligible at $\sqrt{s}\gtrsim$ 10$^{4}$
GeV  while in the U-matrix model it is at $\sqrt{s}\gtrsim$ 10$^{8}$ GeV. Thus an asymptotic regime comes quite late in energy.

In the Table \ref{tab2} the calculated values of the total, elastic and inelastic cross section are given for the LHC energies. One can see a very good agreement with the results of TOTEM \cite{totem3}, ALICE \cite{alice}, ATLAS \cite{atlas} and CMS \cite{cms}.

More problems are for the U-matrix model rather for the eikonal one, however both models do not strongly contradict to the TOTEM data.

\section{Conclusion}
Answering on the question put in the title we would like to say that in our opinion the obtained results demonstrate an ability of the unitarized eikonal and U-matrix 3P+2O models to describe the data up to LHC energies, however, it seems that for further improvement nonlinear and/or non exponential (for vertices) effects are very important.
Exploring another possibility, namely,  construction of the model that does not explicitly violate unitarity (at least the unitarity bounds on the cross sections) just from the beginning its construction is in a progress, results will be presented later.

\begin{table}[h!]
  \centering
  \caption{Parameters of the eikonal and $U$-matrix models obtained by fitting to the data.
  Parameters $\alpha'_{i}$ and $B_{i}$  are given in GeV$^{-2}$, other parameters are dimensionless. Errors are taken from the MINUIT output. $\chi^{2}/dof=\chi^{2}/(N_{p}-N_{par})$  where $N_{p}$ is the number  of experimental points in the fit, $N_{par}$ is the number of free parameters in the model.}
  \label{tab1}
  \medskip
{\small 
 \begin{tabular}{|l|c|c||c|c|}
\hline
 & \multicolumn{2}{|c||}{Eikonal model} &
\multicolumn{2}{|c|}{U-matrix model}\\
\cline{2-5}
 Parameters & value & error &  value & error\\
\hline  $\alpha_{P_{1}}$   &  1.2209 & 0.0012  &  1.2654  & 0.0029   \\
\hline  $\alpha'_{P_{1}}$  &  0.1171 & 0.0007  &  0.1041  & 0.0005   \\
\hline  $g_{P_{1}}$        &  1.3249 & 0.0196  &  0.1371  & 0.0037   \\
\hline  $B_{P_{1}}$        &  0.3977 & 0.0055  &  0.0  & 0.001   \\
\hline  $\alpha_{P_{2}}$   &  1.1675 & 0.0006 &  1.1208  & 0.0006   \\
\hline  $\alpha'_{P_{2}}$  &  0.3036 & 0.0014  &  0.2324  & 0.0048   \\
\hline  $g_{P_{2}}$        &  8.753 & 0.064  &  26.879  & 0.142   \\
\hline  $B_{P_{2}}$        &  0.4125 & 0.0089 &   5.325  & 0.0455   \\
\hline  $\alpha_{P_{3}}$   &  1.0703 & 0.0005  &  1.089  & 0.001    \\
\hline  $\alpha'_{P_{3}}$  &  0.5912 & 0.0024  &  0.2393  & 0.0017   \\
\hline  $g_{P_{3}}$        &  49.138 & 0.198  &  23.657  & 0.161   \\
\hline  $B_{P_{3}}$        &  1.2046 & 0.0169 &  1.2210  & 0.0114   \\
\hline  $\alpha_{O_{1}}$   &  1.220 & 0.0004  &  1.1889  & 0.0001   \\
\hline  $\alpha'_{O_{1}}$  & 0.0723 & 0.7E-05  &  0.0422  & 0.00004   \\
\hline  $g_{O_{1}}$        & -27.133 & 0.007  & -35.072  & 0.026   \\
\hline  $B_{O_{1}}$        &  0.01384 & 0.00003  &  1.052  &  0.0003   \\
\hline  $\alpha_{O_{2}}$   &  1.2196 & 0.00002  &  1.185  & 0.0001   \\
\hline  $\alpha'_{O_{2}}$  &  0.07222 & 0.5E-05  &  0.04197   &0.00004   \\
\hline  $g_{O_{2}}$        &  27.171 & 0.0067  &  36.969   & 0.027    \\
\hline  $B_{O_{2}}$        &  0.01374 & 0.00003  &  1.0671   &0.0003    \\
\hline $\alpha_{+}$        &  0.69 & fixed  &  0.69  & fixed   \\
\hline $\alpha'_{+}$       &  0.84 & fixed  &  0.84  & fixed    \\
\hline $g_{+}$             &  221.55 & 1.18  &  244.49  & 1.05   \\
\hline $B_{+}$             &  5.460 & 0.175  &  0.0  & 0.004   \\
\hline $\alpha_{-}$        &  0.47 & fixed   & 0.47  & fixed   \\
\hline $\alpha'_{-}$       & 0.93 & fixed   & 0.9   &fixed   \\
\hline $g_{-}$             &  149.17 & 4.08  &  182.98  & 2.85   \\
\hline $B_{-}$             & 7.057 & 1.086  &  1.337  & 0.189   \\
\hline
\rule{0pt}{3ex}$\chi^{2}/dof$ & \multicolumn{2}{|c||}{1.673}&
\multicolumn{2}{|c|}{1.973}\\
\hline
\end{tabular}
}
\end{table}


\begin{table}[h!]
  \centering
  \caption{Cross sections and ratio of real to imaginary part of the forward scattering $pp$ amplitude in the eikonal model 3P+2O defined by the Eqs.~(\ref{eq:QEik}), (\ref{eq:3P+2O-model}).}
  \label{tab2}
  \medskip
{\small
 \begin{tabular}{ccccc}
\hline
$\sqrt{s}\,  (TeV)$  & $\sigma_{tot}$ (mb) & $\sigma_{el}$ (mb) &$\sigma_{inel}$ (mb)  &$\rho$  \\
\hline
7    & 98.97 & 25.53 &73.45        &0.153  \\
\hline
 8    & 101.5 & 26.48 &75.02       &0.152  \\
\hline
13   & 111.14  & 30.18  &  80.96   &0.150  \\
\hline
14  & 112.68  & 30.78  & 81.9     &0.150  \\
\hline
\end{tabular}
}
\end{table}

\end{document}